\documentclass[12pt]{article}
\usepackage{graphicx}

\newcommand\pubnumber{NuPhys2016-Pupilli}
\newcommand\pubdate{\today}

\def\INFNPD{INFN Sezione di Padova, I-35131 Padova, Italy}
\def\Title#1{\begin{center} {\Large #1 } \end{center}}
\def\Author#1{\begin{center}{ \sc #1} \end{center}}
\def\Address#1{\begin{center}{ \it #1} \end{center}}

\newcommand\pubblock{\rightline{\begin{tabular}{l} \pubnumber\\
         \pubdate  \end{tabular}}}
\newenvironment{Abstract}{\begin{quotation}  }{\end{quotation}}
\newenvironment{Presented}{\begin{quotation} \begin{center} 
             PRESENTED AT\end{center}\bigskip 
      \begin{center}\begin{large}}{\end{large}\end{center} \end{quotation}}
\def\Acknowledgements{\bigskip  \bigskip \begin{center} \begin{large}
             \bf ACKNOWLEDGEMENTS \end{large}\end{center}}
\def\beq{\begin{equation}}
\def\eeq#1{\label{#1}\end{equation}}
\def\eeqn{\end{equation}}

\def\beqa{\begin{eqnarray}}
\def\eeqa#1{\label{#1}\end{eqnarray}}
\def\eeqan{\end{eqnarray}}

\let\bar=\overbar

\def\Dslash{\not{\hbox{\kern-4pt $D$}}}
\def\dslash{\not{\hbox{\kern-2pt $\del$}}}

\def\msb{{\bar{\ssstyle M \kern -1pt S}}}

\begin{document}
\begin{titlepage}
\pubblock   %%%%% decommentare per inserire data e numero pubblicazione in alto a destra

\vfill
\Title{The ERC ENUBET Project: high precision neutrino flux measurements in conventional neutrino beams}
\vfill
\Author{Fabio Pupilli, on behalf of the ENUBET collaboration}
\Address{\INFNPD}
\vfill
\begin{Abstract}
The challenges of precision neutrino physics require measurements of
absolute neutrino
cross sections at the GeV scale with exquisite (1\%) precision. This
precision is presently
limited by the uncertainties on neutrino flux at the source; their reduction
by one order of magnitude can be achieved monitoring the positron
production in the decay
tunnel originating from the $K_{e3}$ decays of charged kaons in a sign
and momentum selected
narrow band beam. This novel technique enables the measurement of the
most relevant
cross sections for CP violation ($\nu_e$ and $\overline{\nu}_e$) with a precision of
1\% and requires a special
instrumented beam-line. Such non-conventional beam-line will be
developed in the framework
of the ENUBET Horizon-2020 Consolidator Grant, recently approved by
the European Research
Council. The project, the first experimental results
on ultra-compact calorimeters
that can be embedded in the instrumented decay tunnel and the advances
on the simulation of the
beamline are presented. We also discuss the detector and accelerator activities
that are planned in 2016-2021.
\end{Abstract}
\vfill
\begin{Presented}
NuPhys2016, Prospects in Neutrino Physics

Barbican Centre, London, UK,  December 12--14, 2016
\end{Presented}
\vfill
\end{titlepage}
\def\thefootnote{\fnsymbol{footnote}}
\setcounter{footnote}{0}

\section{Conventional and monitored neutrino beams}
In conventional beamlines the decay tunnel is a passive region and the knowledge of the neutrino flux relies on ab-initio
simulations that take into account the proton-target interactions, the reinteraction of secondaries, their tracking and decay. 
Despite the use of hadro-production data from dedicated experiments and of ancillary measurements (proton intensity, horn
currents, beam-target misalignment etc.), the precision on the flux prediction is usually limited to 7-10\%.

A very precise measurement of the $\nu_e$ ($\overline{\nu}_e$) flux can be achieved by directly monitoring in an instrumented
decay tunnel the production of large angle positrons (electrons) from 
$K_{e3}$ decays ($K^{+(-)} \rightarrow \pi^0 e^{+(-)} \nu_e (\overline{\nu}_e)$)
in a sign and momentum-selected narrow band beam \cite{proposal}. The $e^{+}$ rate gives a direct estimate of the $\nu_e$ flux
that is not affected by beam related systematics arising from the number of PoT, the hadro-production cross sections and the
secondary meson focusing efficiency of the beamline, and this method could reduce the uncertainty on the $\nu$ flux normalization
down to 1\%.
%
% The positron rate is monitored in real time but,
% unlike “tagged $\nu$ beams” proposed since the 60’s, leptons are not associated to the observed $\nu$ on an
% event-by-event basis using timing coincidences. The required time resolutions for this application
% are of $\sim$~10 ns thus well within reach of current technologies.

\section{The ENUBET project}

The ENUBET (\textit{Enhanced NeUtrino BEams from kaon Tagging}) project is intended to demonstrate the feasibility of the 
monitored beam approach, by designing and constructing
a detector able to identify positrons from $K_{e3}$ decays in the harsh environment of a neutrino beam decay tunnel \cite{EoI}. 
It will also study the accelerator issues and the the precise layout of the kaon/pion focusing and transport system.
The project has been approved by the ERC (CoG, P.I. A. Longhin, Host Institution INFN) for a five year duration (since 1 June 2016) and
a 2.0 MEUR budget. A controlled neutrino source, like the one proposed by ENUBET, could be exploited by future experiments aiming at
$\mathcal{O}(1\%)$ precision in the electron neutrino cross section measurement. It could be also exploited
in a phase-II sterile neutrino search, especially in case of a positive signal from the upcoming short baseline experiments. 
Finally, ENUBET intends to set the first milestone towards a ``time-tagged neutrino beam'', where the $\nu$
at the detector is time-correlated with the produced $e^{+}$ in the decay tunnel.

%%%%%%%%%%%%%%%%%%%%%%%%%%%%%%%%%%%%%%%%%%%%%%%%%%%%%%%%%%%%%%%%%%%%%%%%%
%%
% %   use this format to include an .eps figure into your paper
% %
\begin{figure}[htb]
\centering
\includegraphics[height=4.9 cm]{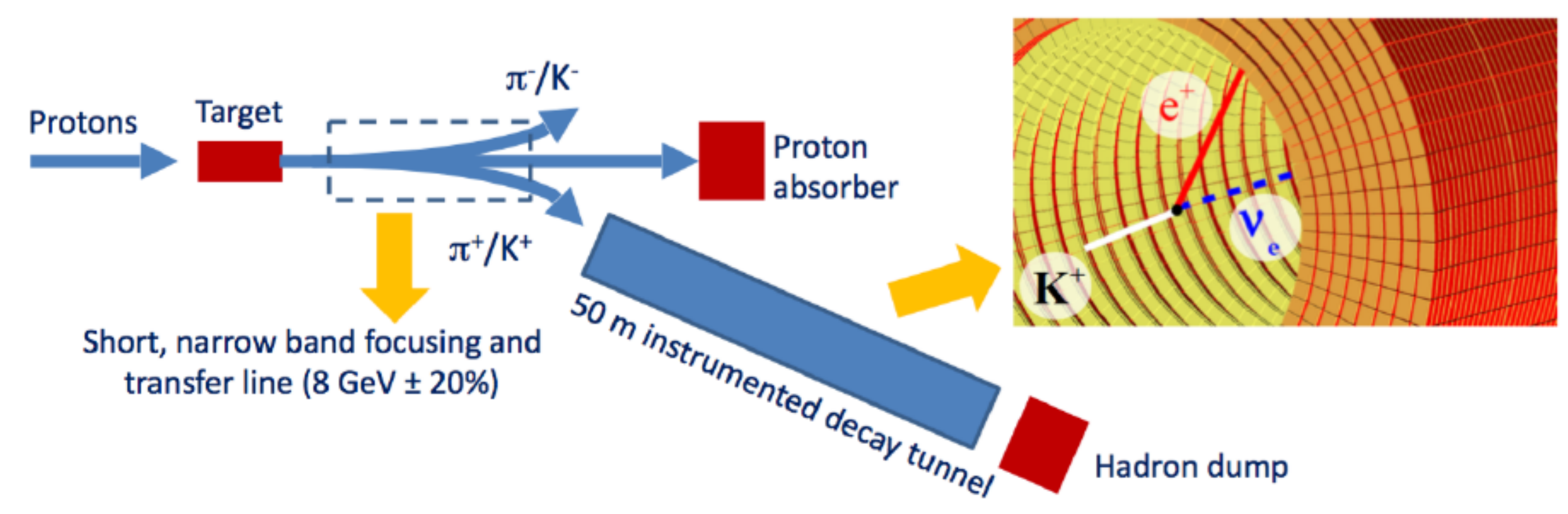}
\caption{The ENUBET beamline. On the right a section of the decay tunnel instrumented with arrays of calorimeter modules
(UCM, in orange) and with the photon veto (in yellow) is also shown.}
\label{fig:enubet}
\end{figure}
%%%%%%%%%%%%%%%%%%%%%%%%%%%%%%%%%%%%%%%%%%%%%%%%%%%%%%%%%%%%%%%%%%%%%%%%%%%

The ENUBET reference design (Figure~\ref{fig:enubet}) foresees a secondary beam with an average momentum of 8.5 GeV and a
$\pm$20\% momentum bite.
The tagging detector is a hollow cylinder surrounding a fraction of the decay tunnel and it is instrumented with a
%longitudinally segmented 
calorimeter for e/$\pi$ separation and a photon veto to separate electromagnetic showers from $\pi^0$ decays
or prompt $e^+$ from $K_{e3}$.

The choice of a high secondary energy and of a short tunnel (50 m) reduces the $\nu_{e}$ content of the beam coming from
muon decays in flight with respect to the ones from $K_{e3}$ and enhances the $\nu_{e}/\nu_{\mu}$ ratio.
Furthermore the resulting positron energy is large enough for an efficient identification through calorimetric techniques and
the produced neutrino spectrum matches the one of interest for future long baseline experiments.
The emittance of the secondary beam (few mrad over 10$\times$10 cm$^2$) and the radius of the tunnel are optimized in order to
prevent undecayed secondaries and muons from pion decay from hitting the calorimeter.

Studies in the preparatory phase of ENUBET have shown that, employing a 500 t neutrino detector (e.g. ICARUS at Fermilab
or ProtoDune-SP/DP at CERN) located 100 m from the entrance 
of the decay tunnel and a 30 GeV (450 GeV) proton driver, 5$\times$10$^{20}$ (5$\times$10$^{19}$)
PoT would be required to provide a sample of 10$^4$ tagged $\nu_{e}^{CC}$ interactions.
The tagging detector can be safely operated in terms of pile-up, dose, etc. if local particle rates are below $\sim$1 MHz/cm$^2$;
this implies that the proton ex\-traction length cannot be shorter than 1 ms. On the other side, longer extractions significantly exceeding
10 ms are disfavoured if secondary focusing is achieved by magnetic horns. The use of a very efficient focusing system based
on DC operated magnets can overcome this limitation and can offer the possibility of reducing particle rates to the level needed
to match current detector time resolutions for the operation of the facility also in ``time-tagged'' mode. Within ENUBET, proton
extraction schemes compatible with accelerators at CERN, J-PARC and Fermilab will be investigated.

\section{Detector prototyping and simulation}

\begin{figure}[htb]
\centering
\begin{minipage}[t]{0.45\textwidth}
\includegraphics[height=4.6 cm]{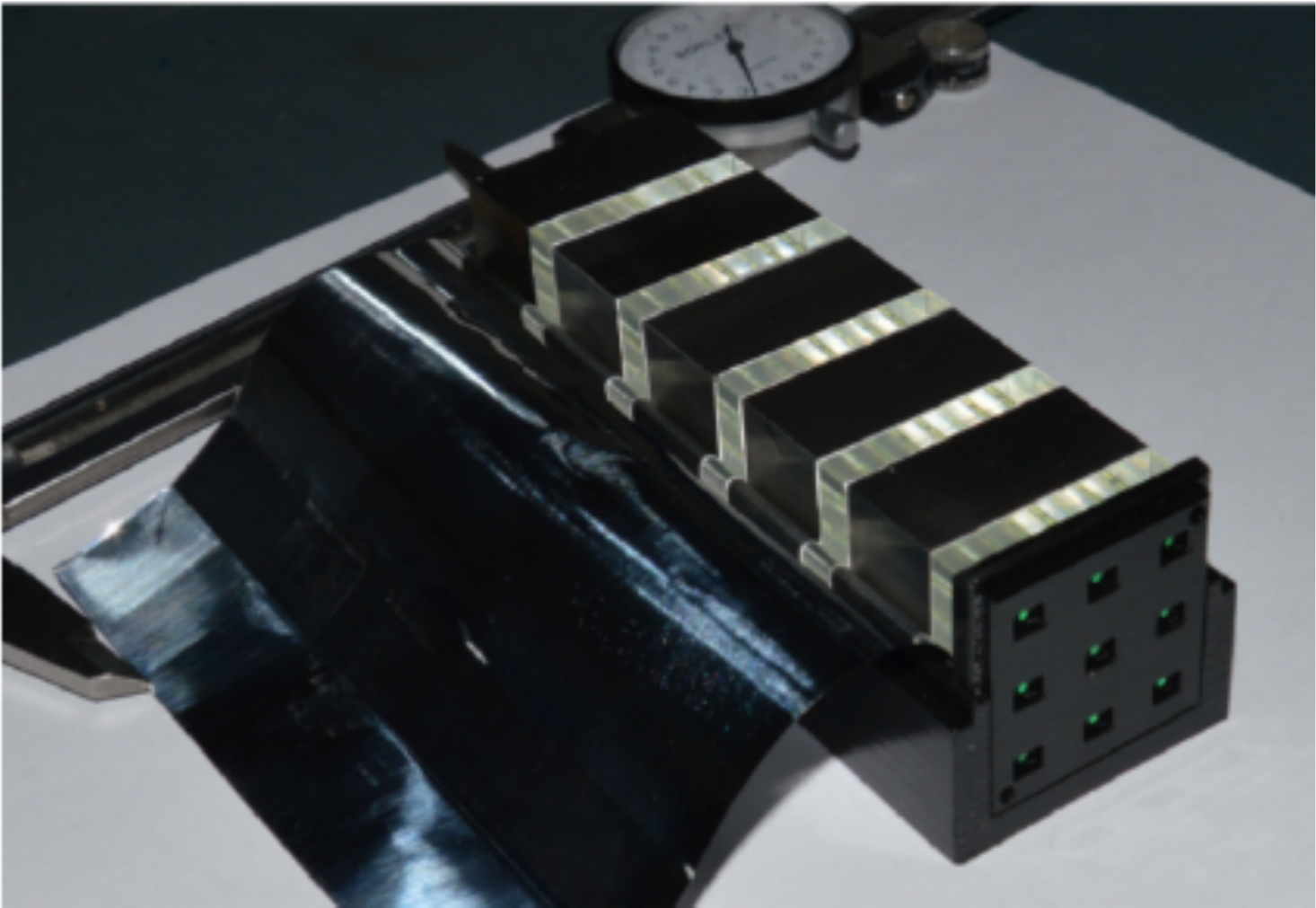}
\end{minipage}
\hspace{10mm}
\begin{minipage}[t]{0.45\textwidth}
\includegraphics[height=4.8 cm]{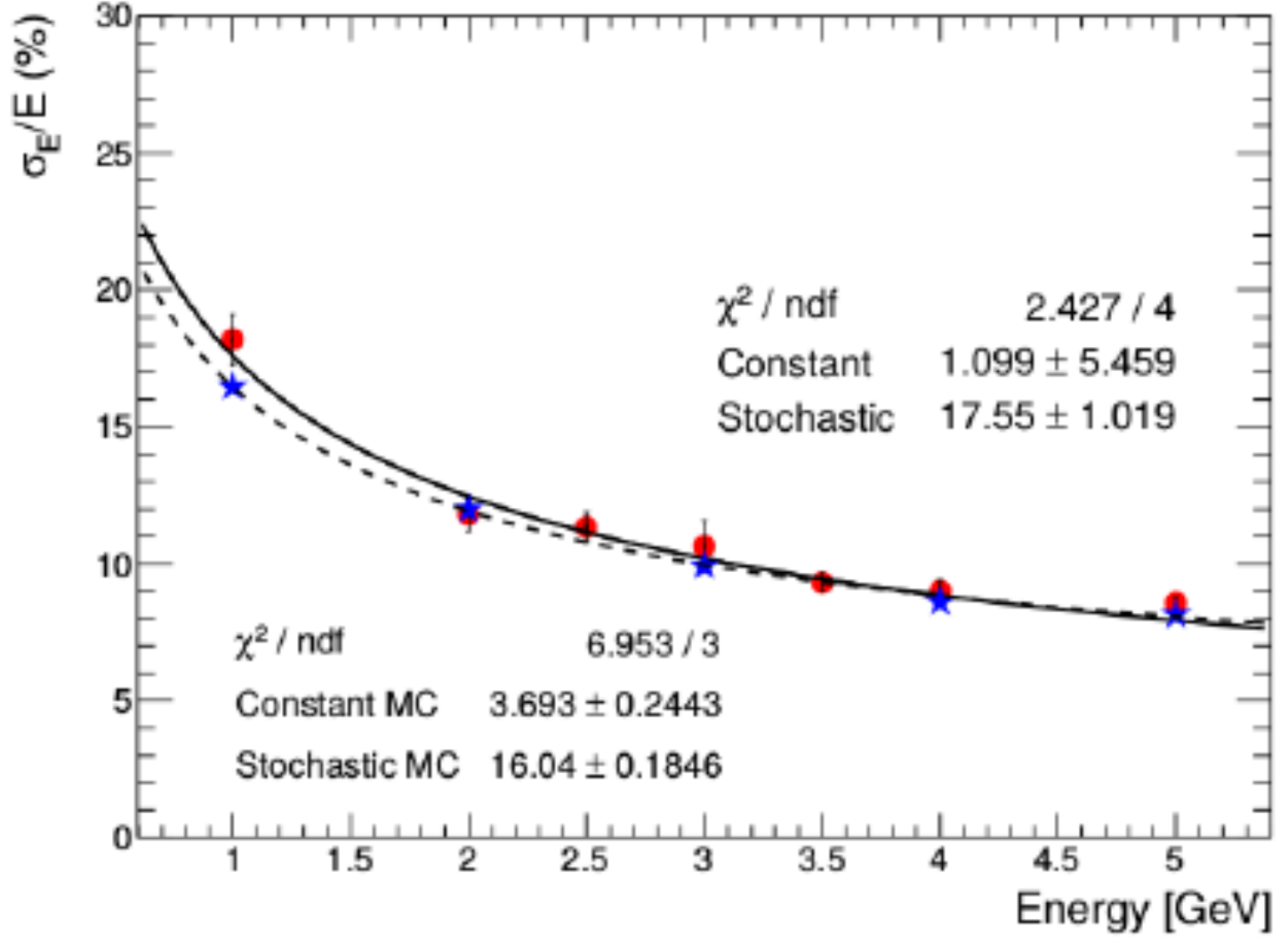}
\end{minipage}
% \caption{(Left) A single UCM. (Right) Electron energy resolution versus beam energy for data obtained in the CERN-PS T9
% exposure (red dots, fit parameters in the top inset), compared to MC simulation (blue stars, fit parameters in the bottom inset)
\caption{(Left) A single UCM. (Right) Electron energy resolution versus beam energy for data obtained in the CERN-PS T9
exposure (red dots), compared to MC simulation (blue stars) \cite{TB2016}.}
\label{fig:UCM}
\end{figure}

In order to cope with the needs of a high $e/\pi$ separation capability
and of a radiation hard, fast-responsive and cost-effective setup, the choice of the tunnel instrumentation
has fallen on a shashlik calorimeter with longitudinal segmentation. The basic unit of the calorimeter, the Ultra-Compact Module (UCM),
is made of five, 15~mm thick, iron layers interleaved by 5~mm thick plastic scintillator tiles. The total length of the module
(10 cm) corresponds to 4.3 $X_0$ and its transverse size is of 3$\times$3 cm$^2$. Nine wavelength shifting fibers crossing the
UCM are connected directly to 1 mm$^2$ SiPMs through a plastic holder (Figure~\ref{fig:UCM}, left).
SiPMs are hosted on a PCB and the output signals are summed and routed toward the front-end electronics by copper-kapton lines.
Unlike conventional
shashlik calorimeters, this scheme avoids the occurrence of large passive regions usually needed to bundle the fibers and route
them to a common photo-sensor, thus greatly improving the homogeneity in the longitudinal sampling.

UCM prototypes were developed by the INFN SCENTT collaboration;
they were tested with cosmic rays and characterized with charged particles in the 1-5 GeV range at the
CERN-PS East Area facility \cite{TB2015,TB2016}. The results have proven the linearity of the calorimeter response
in the considered energy range without saturation up to 4~GeV and an energy resolution of 18\% at 1~GeV 
(Figure~\ref{fig:UCM}, right), well within the requirements of 25\%/$\sqrt E$ for an efficient e/$\pi$ separation.

\begin{figure}[htb]
\centering
\includegraphics[height=5.8 cm]{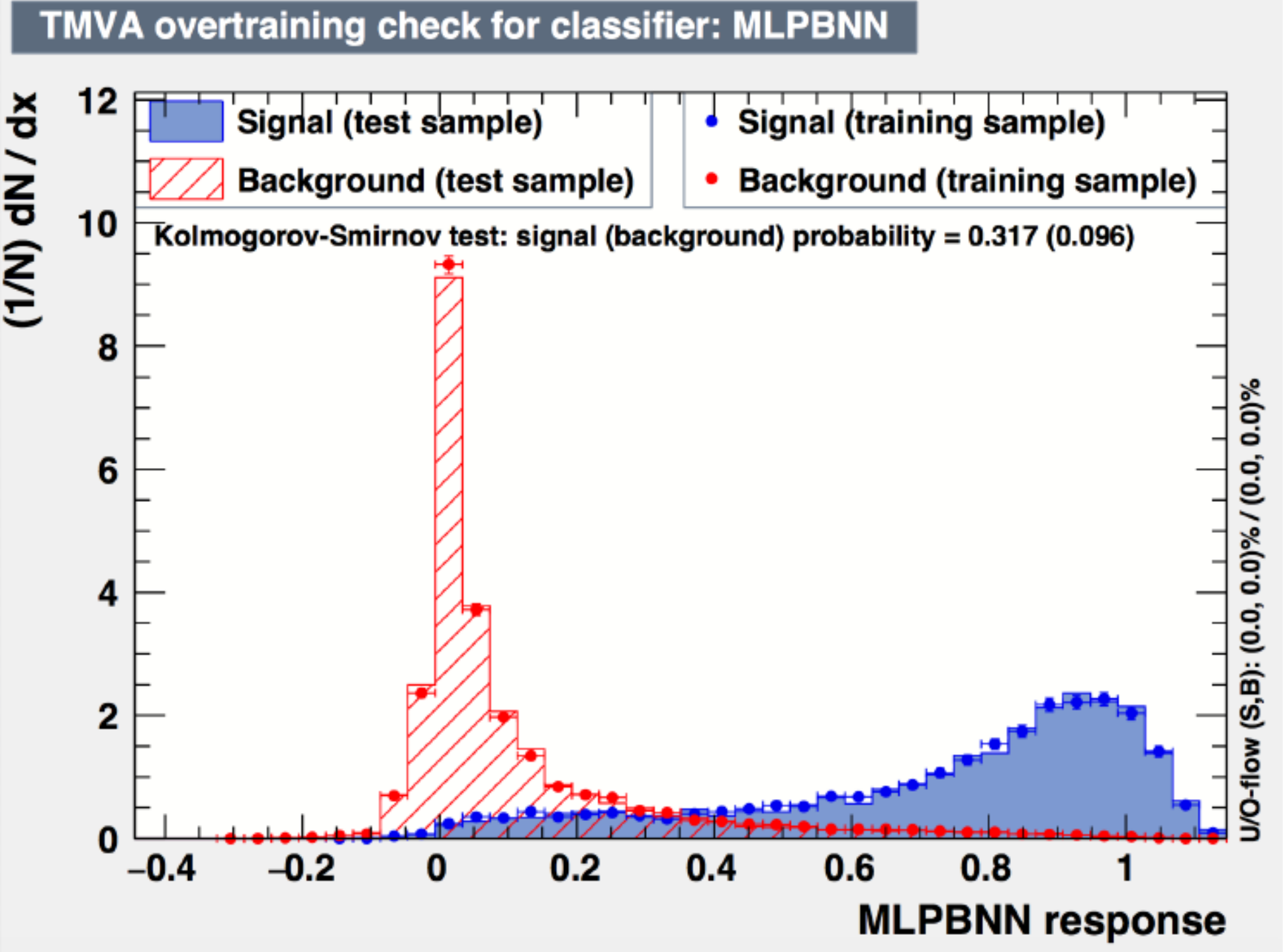}
\caption{TMVA \cite{tmva} neural network response for $\pi^{\pm}$ (red) and $e^+$ (blue).}
\label{fig:tmva}
\end{figure}

The performances of the full instrumented tunnel were simulated with GEANT4~\cite{geant}. The particle identification
algorithms employed rely on the pattern of energy deposit along the UCMs through a multivariate approach based on a neural
network (Figure~\ref{fig:tmva}); the information from hits in the photon veto is used for the $\pi^0$ rejection
\cite{analysis}. For a positron tagging efficiency of 49\%, the $\pi^{\pm}$ and $\pi^0$ mis-identification probabilities are
2.9\% and 1.2\% respectively, confirming that the UCM technique is appropriate for the ENUBET needs.

In november 2016 a module composed of 56 UCMs in 7 longitudinal layers ($\sim$30 $X_0$), complemented by an outer module
acting as energy catcher, have been exposed to electron and pions with various
incidence angles at the CERN-PS; the analysis of these data, currently on going, will allow
to test the e/$\pi$ performance under realistic conditions. 
Further prototype exposures to charged particles (CERN) and neutrons (INFN-LNL CN) are foreseen during
2017-2018, 
in order to also assess the recovery time and the fulfillment of radiation hardness requirements. 
R\&D prototypes of the photon veto will be $\gamma$-irradiated at the INFN-LNF BTF.
Finally a full demonstrator (3 m in length, 180$^\circ$ coverage) will be assembled and tested at CERN.

\Acknowledgements
This project has received funding from the European Research Council (ERC) under the European Union's
Horizon 2020 research and innovation programme (grant agreement No 681647).

\end{document}